\documentclass[11pt]{article}
\usepackage{sectsty}
\usepackage{graphicx}
\usepackage[super,sort&compress,comma]{natbib} 
\usepackage{hyperref}
\usepackage{amsmath}
\usepackage[english]{babel}
\usepackage{epstopdf}%This line makes .eps figures into .pdf - please comment out if not required.
\usepackage[version=3]{mhchem}
\usepackage[left=1.5cm, right=1.5cm, top=1.785cm, bottom=2.0cm]{geometry}
\usepackage{amssymb}
\usepackage{balance}
\usepackage[normalem]{ulem}
\makeatletter
\let\@fnsymbol\@arabic
\makeatother

\title{\textbf{Shear thickening inside elastic open-cell foams \\
under dynamic compression}}
\author{Samantha M. Livermore\thanks{James Franck Institute and Department of Physics, The University of Chicago, Chicago, Illinois 60637, USA}, Alice Pelosse{$^{1}$}, Michael van der Naald{$^{1}$},  Hojin Kim{$^{1,}$}\thanks{Pritzker School of Molecular Engineering, The University of Chicago, Chicago, Illinois 60637, USA} \\
Severine Atis{$^{1,}$}\thanks{Institut Prime, CNRS-Université de Poitiers-ISAE ENSMA, Poitiers, France}, and Heinrich M. Jaeger{$^{1}$}}
\date{\today}

\begin{document}
\maketitle

\section{Abstract}
We measure the response of open-cell polyurethane foams filled with a dense suspension of fumed silica particles in polyethylene glycol at compression speeds spanning several orders of magnitude.
The gradual compressive stress increase of the composite material indicates the existence of shear rate gradients in the interstitial suspension caused by wide distributions in pore sizes in the disordered foam network.
The energy dissipated during compression scales with an effective internal shear rate, allowing for the collapse of three data sets for different pore-size foams.
When scaled by this effective shear rate, the most pronounced energy increase coincides with the effective shear rate corresponding to the onset of shear thickening in our bulk suspension.
Optical measurements of the radial deformation of the foam network and of the suspension flow under compression provide additional insight into the interaction between shear thickening fluid and foam.
This optical data, combined with a simple model of a spring submerged in viscous flow, illustrates the dynamic interaction of viscous drag with foam elasticity as a function of compression rate, and identifies the foam pore size distribution as a critically important model parameter.
Taken together, the stress measurements, dissipated energy, and relative motion of the fluid and the foam can be rationalized by knowing the pore size distribution and the average pore size of the foam. 

\pagebreak

\section{Introduction}
Concentrated suspensions of rigid particles exhibiting discontinuous shear thickening (DST) can transform from fluid-like to solid-like under shear\cite{Barnes_1989,Morris_2020,Mari_Seto_Morris_Denn_2015,brown2014shear}.
The highly dissipative nature of this transition can be harnessed in a variety of applications\cite{Lin_Ness_Cates_Sun_Cohen_2016,Niu_Ramaswamy_Ness_Shetty_Cohen_2020,Richards_Hodgson_ONeill_DeRosa_Poon_2024}.
% This state transition, in response to applied force, is highly dissipative and can either be an obstacle or a purposeful feature, depending on the applications \cite{Lin_2015}.
In particular, DST has led to a variety of composite materials, designed for low-velocity impact mitigation \cite{Wang_Zhang_Gao_Gao_He_2024}. 
Specifically, this includes fabrics\cite{Decker_Halbach_Nam_Wagner_Wetzel_2007,Wagner_Armor,Fehrenbach_Hall_Gibbon_Smith_Amiri_Ulven_2023,Liu_Zhang_Liu_Cao_Wang_Bai_Sang_Xuan_Jiang_Gong_2019} and foams\cite{parisi_indentation_2023,Caglayan_Osken_Ataalp_Turkmen_Cebeci_2020,Fischer_Braun_Bourban_Michaud_Plummer_Månson_2006,soutrenon_impact_2014,dawson_dynamic_2009,Sheikhi_Gürgen_2022,Zhao_Wu_Lu_Lin_Jiang_2022,haris_effectiveness_2018} saturated with shear thickening fluid, and foams made of a shear thickening gel\cite{Fan_Xue_Zhu_Zhang_Li_Chen_Huang_Fu_2022,Xiaoke_Kejing_Qianqian_Kun_2019}. 
These materials are flexible at low applied stress but become rigid under sudden impact, making them suitable for impact protection as well as damping applications\cite{Srivastava_Majumdar_Butola_2012,Fischer_Bennani_Michaud_Jacquelin_Månson_2010,Iyer_Vedad-Ghavami_Lee_Liger_Kavehpour_Candler_2013}. 
  
Of the aforementioned composite materials, some of the most remarkable behavior arises in open-cell foams saturated with shear thickening fluid (STF).
An STF-foam composite can generate a dramatically enhanced stress response under conditions where the bulk suspension alone would not exhibit significant thickening and an open-cell foam offers only weak resistance to compression.
Prior work has associated this enhancement with an increased effective shear rate $\dot{\gamma}_\mathrm{eff}$ experienced by the STF as it is forced to flow through narrow channels and constrictions inside the foam\cite{soutrenon_impact_2014,dawson_dynamic_2009}:
\begin{equation}
    \dot{\gamma}_\mathrm{eff}= \frac{\mathrm{v_c}}{d_\mathrm{foam}},
    \label{eqn:eqn1}
\end{equation}
\noindent where $\mathrm{v_c}$ is the rate of compression of the foam and $d_\mathrm{foam}$ is some length scale characterizing the internal foam geometry. 
When $\dot{\gamma}_\mathrm{eff}$ reaches the critical shear rate for the onset of shear thickening in the neat STF, the STF-foam composite's resistance to compression increases strongly. 
While this captures the qualitative behavior of the material, it remains difficult to link this effective shear rate to the foam geometry.
Furthermore, this description only considers how the foam affects the STF and neglects the opposite interaction, namely how the STF changes the behavior of the foam.
This feedback becomes important during shear thickening, where the rapidly increasing viscosity of the STF can generate significantly enhanced drag on the foam.  

\begin{figure*}[!t]
\centering
  \includegraphics[width=\linewidth]{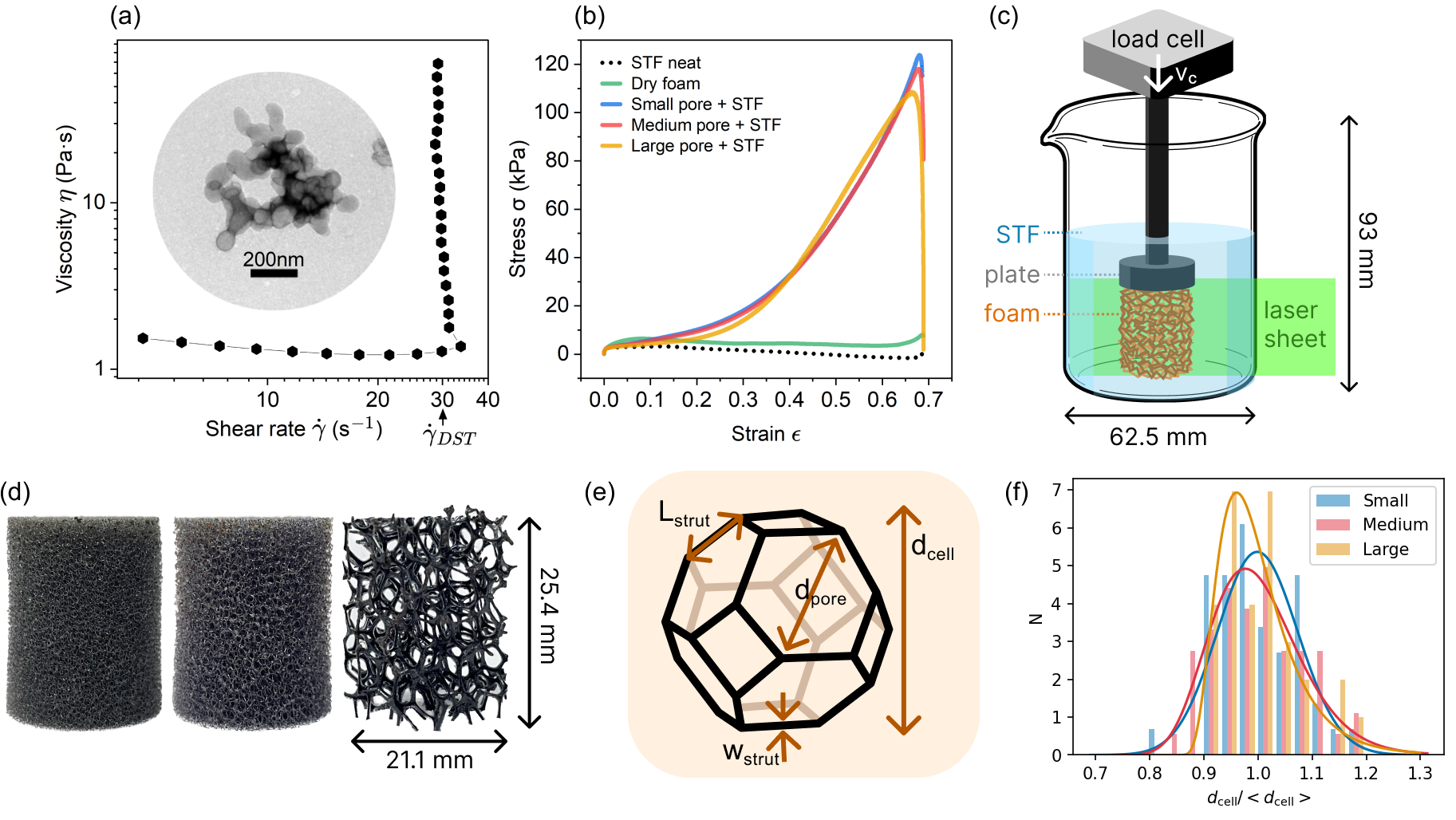}
  \caption{
  (a) Steady-state rheology for the neat shear thickening fluid (STF), a fumed silica and PEG suspension at volume fraction $\phi_{V}=30\%$.
  Inset: Transmission electron microscope image of a fumed silica particle.
  (b) Stress-strain responses of the neat STF, neat (dry) foam, and the three composite STF-foam materials with different pore sizes, compressed at speed $\mathrm{v_c}=150$~mm/s.
  (c) Schematic of the experimental apparatus. 
  A cylinder of foam is submerged in a volume of STF and centered underneath a compression plate connected to a load cell, which moves downward at speed $\mathrm{v_c}$ during testing. The compression plate is 25.4 mm in diameter and 13.4 mm tall. 
  (d) Images of the three different foam types with small, medium, and large pores (left to right).
  (e) Schematic of an idealized foam cell.
  (f) Normalized distributions of the cell diameters for the three foam varieties. 
  Cell sizes are normalized by their mean value $\langle d_\text{cell}\rangle$.
  }
  \label{fgr:fig1}
\end{figure*}

Here we investigate both directions of this dynamic interaction in open-cell foams submerged in a bath of STF and then compressed.
For this purpose, we choose a suspension comprised of fumed silica nanoparticles dispersed at 30\% solid volume fraction $\phi_\mathrm{V}$ in polyethylene glycol (PEG).
Up to a critical shear rate $\dot{\gamma}_\mathrm{DST} \approx 30$~s$^{-1}$, this suspension is Newtonian with nearly constant viscosity $\eta$, at which point the viscosity jumps up discontinuously by over an order of magnitude (Fig.~\ref{fgr:fig1}a); this is known as discontinuous shear thickening (DST)\cite{Brown_Jaeger_2009,Bourrianne_Niggel_Polly_Divoux_McKinley_2022}.
At vertical loading speeds of $\mathrm{v_c}=$150~mm/s, neither the neat suspension nor the dry foam produces significant stress on the load cell, while the STF-foam composites generate stresses $\sigma$ that exceed 100~kPa as a function of axial compressive strain $\epsilon$ (Fig.~\ref{fgr:fig1}b).
Figure ~\ref{fgr:fig1}c shows the experimental setup used for compression tests of the composite STF-foam material.
We perform compression tests with the same suspension on three open-cell polyurethane foams with different average cell sizes (Figs.~\ref{fgr:fig1}d, e and Table 1).
Each individual foam network contains a wide distribution of cell sizes about the mean $\langle d_\mathrm{cell}\rangle$ (Fig.~\ref{fgr:fig1}f).
The smooth increase of the data in Fig.~\ref{fgr:fig1}b, and the fact that the stress curves for $\mathrm{v_c}=150$~mm/s display similar behavior (despite their average cell sizes differing by almost an order of magnitude) highlights the difficulty of associating a single length scale $d_\mathrm{foam}$ in eqn.~(\ref{eqn:eqn1}) with the stress-strain response of the composite material.
Nevertheless, we find that a scaling with characteristic $\dot{\gamma}_\mathrm{eff}$ emerges if instead we consider the energy dissipated during compression, i.e., the area under stress-strain curves as in Fig.~\ref{fgr:fig1}b.
We then show that this energy scales with $\mathrm{v_c}/d_\mathrm{foam}$ for different foam types, and undergoes a pronounced increase centered around $\mathrm{v_c}/d_\mathrm{foam} = \dot{\gamma}_\mathrm{DST}$, if we take the foam length scale to be the average pore diameter of the foam pore size distribution.

The optical transparency of our suspension enables us to study the effect of the STF on the foam by tracking the deformation of the foam network under the influence of viscous flow.
A dry open-cell foam compresses by successively collapsing layers of cells perpendicular to the direction of compression, causing the foam to maintain a near-constant radius under compression.
However, when the foam is instead filled with a viscous fluid that is forced to flow out radially during compression, the foam struts experience viscous drag forces that can cause the foam to bulge out radially.
Comparing the speed of radial foam expansion to the radial fluid speed just outside the foam reveals how viscous drag affects the foam along with signs of viscosity changes in the STF. 
Finally, analyzing these data with a simple model makes it possible to relate the measured speed of radial foam expansion to both the average pore size during compression and the width of the pore size distribution. 

\section{Materials and methods}

\subsection{Dense suspension preparation and rheology}

We use a suspension of fumed silica (Aerosil OX-50, Evonik/Palmer Holland, average particle size $\sim$500~nm, dry density 2.2~g/cm$^3$, see Fig.~\ref{fgr:fig1}a inset) and polyethylene glycol (PEG, Sigma-Aldrich, $M_W$=200~g/mol, density 1.1~g/cm$^3$, $\eta=50$~mPa$\cdot$s). 
The PEG and fumed silica are combined using a Heidolph high-torque mechanical mixer (ramping from 500 to 3000~rpm), starting with the full volume of PEG and gradually adding the fumed silica. 
Adding the particles first results in a gel-like material that is difficult to mix. 
The mixture is sonicated and mechanically mixed in cycles until a steady rheological state is reached. 
Due to the strong shearing forces from the initially inhomogeneous suspension, active supervision is necessary to prevent the sample container from spinning out or the mixing attachment from grinding against the chuck jaws.

The shear rheology is sensitive to temperature increase during mixing, so suspensions are allowed to rest for several hours before any measurements are taken. 
The entire preparation protocol (mechanical mixing, sonication, resting, testing) typically takes over a week to produce $\sim300$~mL of suspension. 
The suspension is stored under house vacuum to prevent the hydrophilic PEG from absorbing moisture from the air\cite{pereira_combining_2013}. 
To minimize rheological disparities during STF-foam testing, the same large STF batch is used for all compression tests on all three foam types.
Periodic rheological measurements ensure there are no aging effects over time.

All shear thickening suspensions discussed in the following refer to Aerosil OX-50 in PEG-200 with a volume fraction of $\phi_\mathrm{V}=0.30$. The density of these suspensions is $1.43$~g/cm$^3$.
The suspension rheology is measured using a stress-controlled rheometer (Anton Paar MCR301) with a 25~mm parallel-plate geometry and a gap size of 0.5-1~mm. 
Dynamic viscosity is recorded with an initial ascending ramp from 7-1000~Pa, followed by a descending ramp back to 7~Pa, where the upper limit is chosen to prevent sample ejection.
Figure~\ref{fgr:fig1}a is the measured viscosity of the STF as a function of shear rate, showing pronounced DST at a shear rate $\dot{\gamma}_\mathrm{DST} \approx 30$~s$^{-1}$.

\begin{table*}[!t]
\centering
\small
  \caption{\ Foam parameters. 
  }
  \label{foamspecs}
  \begin{tabular*}{\textwidth}{@{\extracolsep{\fill}}lllll}
    \hline
    Foam label & Cell size (mm) & Strut thickness ($\mu$m) & Avg strut length (mm)\\
    \hline
        Large pore & 6.1 $\pm$ 0.4 & 490 $\pm$ 83 & 2.7 $\pm$ 0.9\\
        Medium pore & 1.2 $\pm$ 0.1 & 110 $\pm$ 16 & 0.50 $\pm$ 0.1\\
        Small pore & 0.85 $\pm$ 0.06 & 56 $\pm$ 9.0 & 0.28 $\pm$ 0.07\\
    \hline
    \label{table1}
  \end{tabular*}
\end{table*}
\subsection{Foam preparation and characterization}
 
We test three grades of ester-based polyurethane (PU) open-cell foam (U.S. Plastics Corp.), with small, medium, and large pores, shown in Fig.~\ref{fgr:fig1}d.
For average cell size, strut length, and strut thickness see Table~\ref{table1}.
The data in this table are from optical measurements (DSX-1000 digital microscope, Olympus), assuming all in-focus foam elements are "in-plane" when extracting lengths using image analysis software.
Although the foam cells often are taken as \textit{Kelvin cells} (Fig.~\ref{fgr:fig1}e), they are in fact largely disordered cells with varying numbers of 3,4, and 5-sided faces\cite{zhu_analysis_1997}.
Distributions of the cell diameters $d_\mathrm{cell}$ for all three foams are shown in Fig. \ref{fgr:fig1}f; the slight left skew of the distribution has been discussed previously in the literature\cite{elliott_-situ_2002,Jang_Kraynik_Kyriakides_2008}.  
Analysis of PU open-cell foams of different $d_\mathrm{cell}$ has shown\cite{Mills_2005,Jang_Kraynik_Kyriakides_2008} that the average pore size, i.e., the average size of the various openings in the cell facets, scales with average cell size according to $\langle d_\mathrm{0}\rangle=\beta \langle d_\mathrm{cell}\rangle$, with a factor $\beta$ between $\frac{1}{3}$ and  $\frac{1}{5}$.
This factor depends on the details of the manufacturing protocol and the relative size of the network struts. 
In the following we take $\langle d_\mathrm{0}\rangle=\frac{1}{4} \langle d_\mathrm{cell}\rangle$, which should be appropriate for our medium and small pore foams.

Foam specimens are cut into cylinders (21.1~mm diameter, 25.4~mm height) using a laser cutter (ULTRA X6000, Universal Laser Systems).
The diameter of the cylinder face closest to the laser typically is $\sim 1$~mm smaller than the face furthest from the laser due to heating.

\subsection{STF-filled foam sample preparation}

Individual foam cores are submerged into $\sim$150~mL of well-mixed suspension; this leaves $\sim$2 cm of fluid clearance above the foam,  ensuring that the impactor plate is always submerged in fluid before axial strain is applied (see sketch in Fig. \ref{fgr:fig1}c).
To remove trapped air bubbles, samples are placed under 2 Torr vacuum for 10 minutes prior to compression. 

\subsection{Compression testing protocol}

The stress-strain response of each foam-STF sample under compression is measured using a universal testing instrument (Zwick-Roell Z1.0, equipped with a 1~kN load cell). 
Stress is calculated as the raw force reading divided by the surface area of the foam's top face, the area of which is assumed to remain constant. 
The strain $\epsilon$ is the global axial strain, given by $\epsilon(t) = \frac{h(t)-h_0}{h_0}$, where $h(t)$ is the height at time $t$, and $h_0$ is the original height of the foam. 
The impactor plate and the surface below the foam are coated with sandpaper to prevent the foam from slipping laterally.

The compression protocol includes several steps. 
First, the plate's starting height is adjusted so it is submerged in the suspension to minimize bubble formation and other possible suspension-air interfacial effects. 
The plate then moves downward onto the foam until it reaches a pre-stress of 0.2~N. 
Next the STF-foam sample undergoes two conditioning cycles at 3~mm/s to a strain of 70\% (corresponding to a height change of 17.5~mm) in order to standardize the shear history of the sample\footnote{Foam samples submerged in silicone oil, used for comparison with foam-STF samples, do not undergo conditioning due to the risk of slipping.}. 
This depth was chosen because PU foams are resilient against repeated compression up to 75\%\cite{hager_fatigue_1992}. 

After conditioning, the plate pauses above the sample for 3~s before commencing the measurement phase at a preset impact speed, which we vary from 0.1~mm/s to 150~mm/s, corresponding to strain rates from 0.004~s$^{-1}$ to 5.91~s$^{-1}$ as defined by $\mathrm{v_c}/h_\mathrm{0}$. 
Although the compression tests are speed-controlled, there are brief acceleration transients at the beginning and the end of the pre-set travel distance  (17.5~mm) while the system is ramping up to the targeted speed and during stopping.   
We therefore limit the detailed analysis of our motion tracking data to the strain range $0.2 < \epsilon < 0.4$ to  avoid acceleration effects even at the fastest compression speeds.

\subsection{Particle image velocimetry protocol}

Compressions are filmed with a Phantom high-speed camera (V12 and VEO models, 1200$\times$708~px image size). 
For particle image velocimetry (PIV), filming rates of 50-4000~fps are used to accurately correlate particle motion between frames.

To map the flow of the STF at the boundaries of the foam, fluorescent red polyethylene microspheres (Cospheric, density 1.2~g/cm$^3$) with a diameter of 75-90~$\mu$m are added to the suspension as tracers.
The comparable densities of the tracer particles and the suspension allow the tracers to remain evenly suspended for several hours at a time.
A vertical slice of the suspension is illuminated with a laser sheet of 505-550~nm wavelength (Motovera self-leveling green laser); this plane is perpendicular to the axis of compression and to the camera. 
The laser sheet is about 2~mm thick at close range. 

Examples of  high-speed videos are included as Supplementary Materials.
Videos are analyzed using the openPIV Python package. 
The output of the PIV analysis is a 2D vector field ($\mathrm{v_r}$,$\mathrm{v_z}$) for every frame, describing the flow field outside of the foam at that time step. 
We use an interrogation window size of 32~px in the leading frame, a search area size of 36~px in the following frame, with an overlap of 17~px.
We select these parameters to include several tracer particles in each interrogation window.
In addition to tracking the fluid flow, we also track the radial deformation of the foam edge. 
For each video frame, we fit a spline to the foam edge in order to distinguish it from the surrounding fluid. 
We subdivide the spline into 20 equal length pieces and track these individual segments over time. 
This allows us to compare the foam and fluid motions locally.

\begin{figure*}[!t]
\centering

  \includegraphics[width=\linewidth]{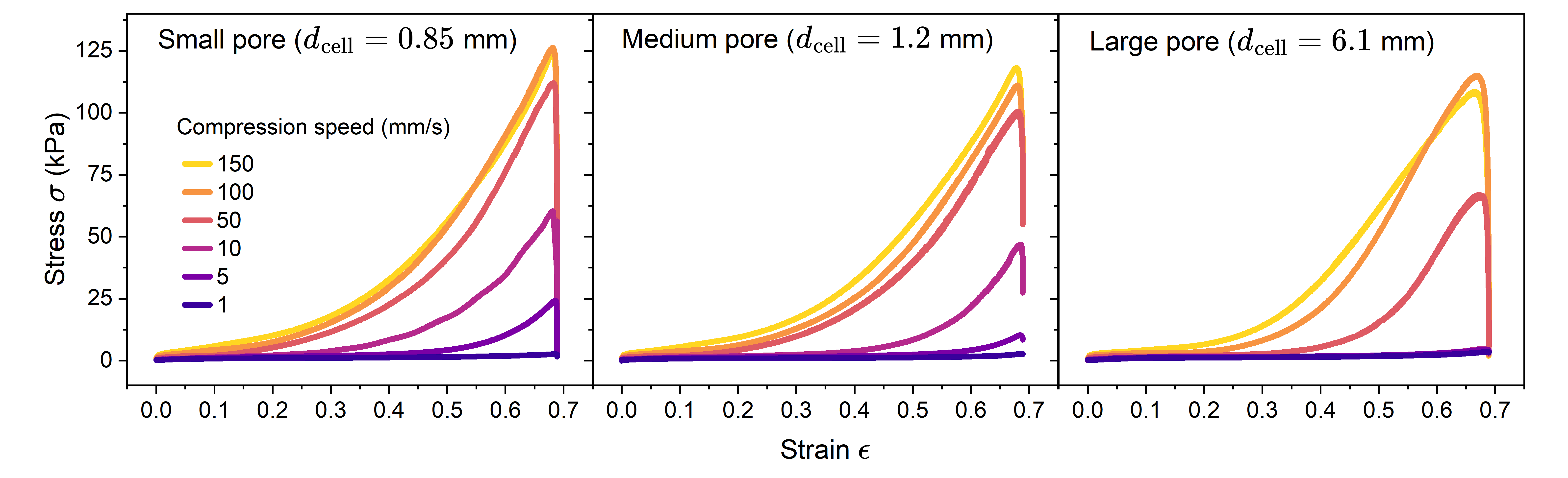}
  \caption{
  Average compressive stress-strain response of the small, medium, and large pore foams, from left to right, for a range of impact speeds. 
  Each curve is the average of at least two compression tests at a given compression speed. 
  The full set of impact speeds is presented in the SI.
   }
  \label{fgr:fig2}
\end{figure*}

\section{Results and discussion}

\subsection{Stress-strain response of STF-foam composites}

For a given foam type, the compressive stress response of the STF-foam composite increases with increasing compression speed $\mathrm{v_c}$, as shown in Fig.~\ref{fgr:fig2}.
Across various pore sizes, larger pores require faster compression to achieve a given stress.
This trend is best illustrated by the $10$~mm/s curve in all panels of Fig.~\ref{fgr:fig2}, where the onset of non-zero stress occurs at larger strains as the foam pore size increases. 
A key feature of Fig.~\ref{fgr:fig2} is the smooth shape of the stress-strain response, which is in stark contrast to the highly discontinuous rheology of the neat STF at the onset of DST (Fig.~\ref{fgr:fig1}a).
Furthermore, we observe a saturation of the stress response:  as previously shown in Fig.~\ref{fgr:fig1}b, at 100~mm/s and 150~mm/s, the curves become effectively independent of foam type and thus pore size.
These observations are consistent with the literature \cite{dawson_dynamic_2009,soutrenon_impact_2014}.
In particular, \citet{dawson_dynamic_2009} find that while the stress response of their STF-foam is highly dependent on pore size at low strain rates, it becomes pore-size independent at shear rates exceeding the shear thickening threshold.

However, these data are clearly incompatible with a single length scale $d_\mathrm{foam}$ characterizing the internal foam geometry in eqn.~(\ref{eqn:eqn1}), even if this length scale is strain dependent.
The measurements indicate that any such length scale must have an inherently wide distribution, which in turn generates a wide range of effective, local shear rates within the foam. 
We can rationalize the smooth stress response and its saturation behavior using the distribution in pore sizes in the foam.
As the foam network is compressed, pores of varying size shrink as a function of strain.
This means that, over the course of a compression, more and more of the STF inside the foam becomes shear thickened as the fraction of pores small enough to trigger DST increases.
Once $\mathrm{v_c}$ is sufficiently large that $\dot{\gamma}_\mathrm{DST}$ is met for even the largest pores from the very beginning of the compression, everything shear thickens at once.
Increasing $\mathrm{v_c}$ further will not change the stress response, leading to the stress saturation we observe in Fig.~\ref{fgr:fig2} for a given foam type.
For $\mathrm{v_c}=$100 and 150~mm/s, the compression speed is then large enough to have triggered DST across even the foam with the largest pores.

\begin{figure}[!t]
\centering
  \includegraphics[width=3in]{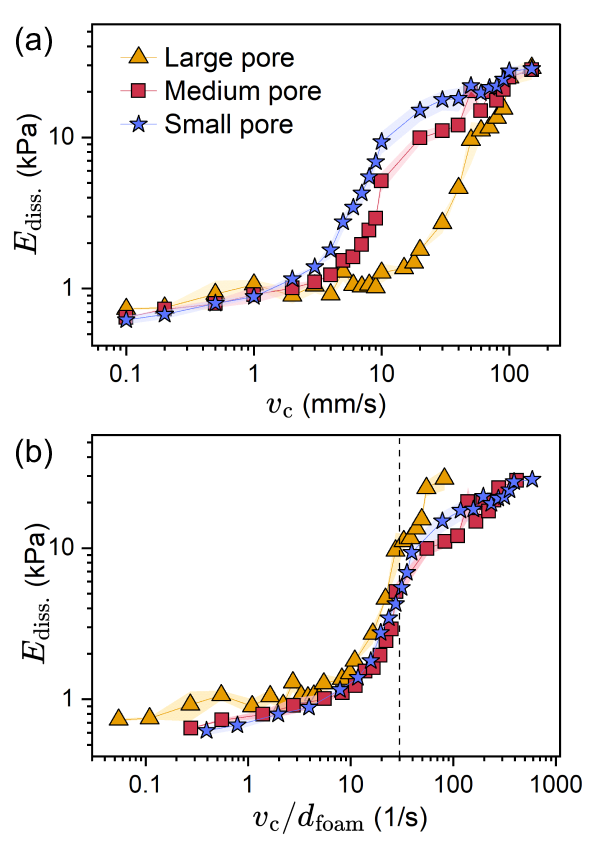}
  \caption{
  (a) Energy dissipated as a function of compression speed $\mathrm{v_c}$ for three pore sizes. 
  (b) Same data as in A but plotted as a function of effective shear rate $\mathrm{v_c}/d_\mathrm{foam}$, where the average pore size $\langle d_0 \rangle$ was taken as the characteristic length scale $d_\mathrm{foam}$, showing collapse of  the three curves.
  The vertical dashed line denotes $\dot{\gamma}_\mathrm{DST}=30$~s$^{-1}$.
  }
  \label{fgr:fig3}
\end{figure}

While linking the rheology of the neat STF to the detailed compression response of the composite material via the distribution of pore sizes is not straightforward for the stress-strain data (Fig.~\ref{fgr:fig1}b and Fig.~\ref{fgr:fig2}), such a link emerges naturally when considering the energy dissipated $E_\mathrm{diss}$ during compression.   
$E_\mathrm{diss}$ is given by the integral of the stress $\sigma$ over a strain range $\epsilon \in (\epsilon_i,\epsilon_f)$,
\begin{equation}
    E_\mathrm{diss} = \int_{\epsilon_i}^{\epsilon_f} \sigma d \epsilon
    \label{eq:toughness}
\end{equation}
\noindent and is shown in Fig.~\ref{fgr:fig3}.  
Here $\epsilon_i=0.2$ and $\epsilon_f=0.65$ for all data and small, medium, and large pore sizes are shown as blue stars, red squares, and gold triangles, respectively.  
For all foam types, as a function of compression speed $E_\mathrm{diss}$ first increases gradually, then exhibits a rapid increase by about an order of magnitude, which shifts to larger $\mathrm{v_c}$ for larger pore sizes, and finally returns to a more gradual increase. 

If the same $E_\mathrm{diss}$ data are instead plotted as a function of an effective shear rate $ \mathrm{v_c}/d_\mathrm{foam}$ they can be collapsed by making $d_\mathrm{foam}$ proportional to $\langle d_\mathrm{cell}\rangle$.
Most interestingly, if we equate the characteristic length scale $d_\mathrm{foam}$ with the mean of the pore size distribution,
$\langle d_0\rangle$, and use that  $\langle d_0\rangle \approx \frac{1}{4}\langle d_\mathrm{cell}\rangle$ in polyurethane open-cell foams,\cite{Mills_2005,Jang_Kraynik_Kyriakides_2008}
we find the data collapse such that the steepest rise in $E_\mathrm{diss}$ occurs right around $\dot{\gamma}_\mathrm{eff} =  \dot{\gamma}_\mathrm{DST} = 30$~s$^{-1}$ for the neat STF; this is shown in Figure~\ref{fgr:fig3}b.
For effective shear rates $ \dot{\gamma}_\mathrm{eff} < 10\,~\text{s}^{-1}$, there is only gentle growth of $E_\mathrm{diss}$ in the composite material. 
This  aligns with the idea that at low $\mathrm{v_c}$, only a small fraction of pores in the foam lead to DST in the suspension, so the overall effective viscosity of the sample remains low.
At intermediate rates $10 < \dot{\gamma}_\mathrm{eff} < 50$~s$^{-1}$, the viscosity of the STF inside the composite is highly sensitive to the pore size distribution, leading to the strong increase in $E_\mathrm{diss}$.
This coincides with a greater fraction of foam pores initiating DST in the suspension, such that the overall effective viscosity of the suspension increases strongly.
Finally, for $\dot{\gamma}_\mathrm{eff} \geq 50$~s$^{-1}$, the reduced slope in $E_\mathrm{diss}$ shows that the majority of pore sizes has now become involved in generating DST. 

Therefore, the shape of the transition between these regimes in $E_\mathrm{diss}$ reflects the shape of the (cumulative) distribution of the pore sizes.
The collapse of the three datasets across the whole transition region thus indicates that despite having different means, their pore size distributions must have nearly the same shape.

\subsection{Velocimetry measurements of foam and suspension}

We leverage the optical transparency of our suspension to monitor the expansion of the compressed foam and the flow of the STF local to the foam boundary as described in \S~2.5. 
We make these measurements for a small pore foam submerged in STF, and also for a foam submerged in a high-viscosity, Newtonian silicone oil ($\eta=$10~Pa$\cdot$s) as a control to distinguish non-Newtonian effects from Newtonian ones.
We determine the relative magnitudes of spring forces from the elastic foam network and the viscous drag forces of the filler fluid using the relative radial motion of the foam boundary and the local fluid.

\begin{figure*}[!b]
    \centering
    \includegraphics[width=\linewidth]{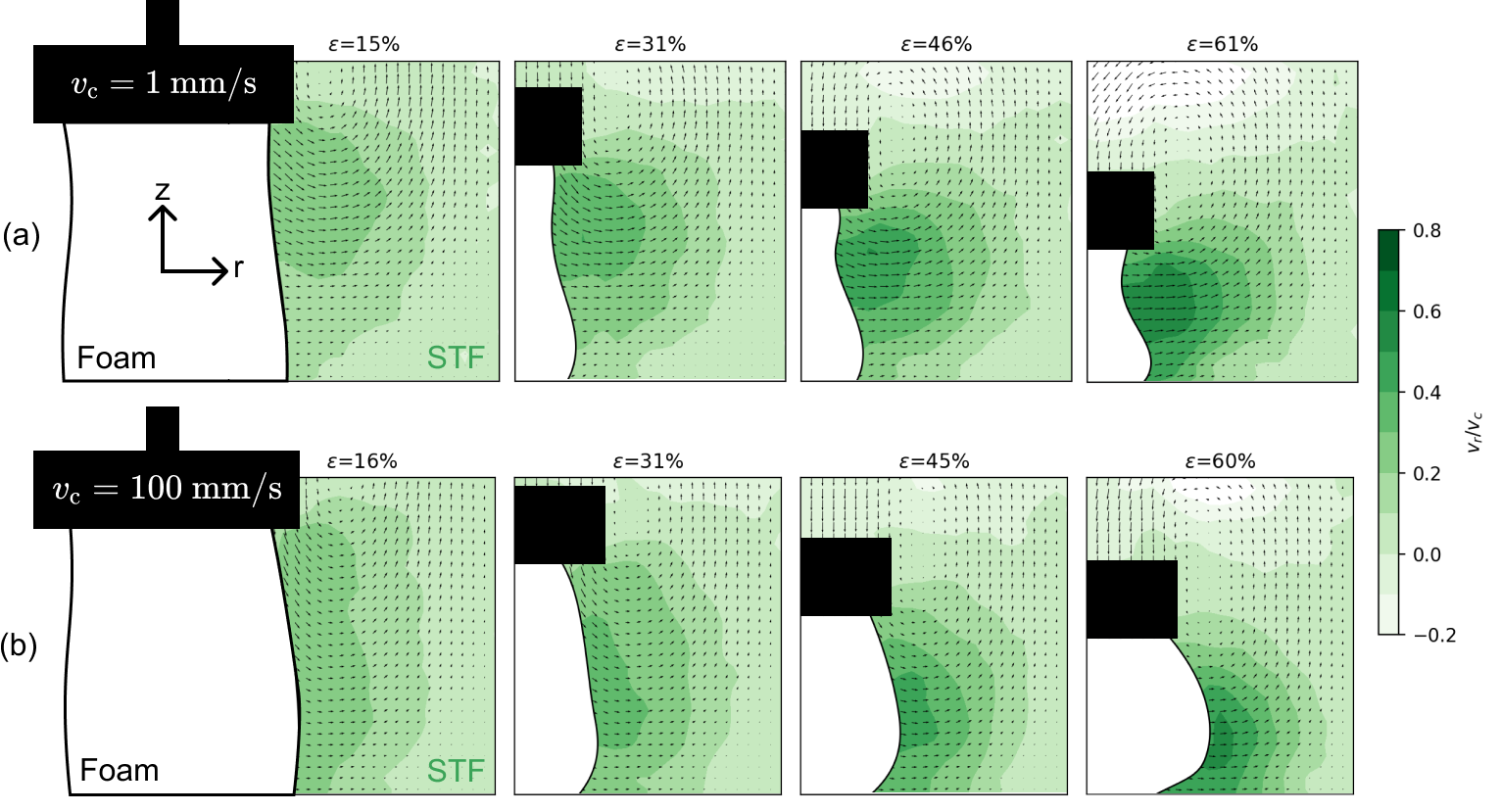}
    \caption{
    Particle image velocimetry (PIV) analysis of the fluid outside the foam for different strains $\epsilon$. 
    The 2D vector field includes both radial and axial velocity components; the contour shading in green denotes the magnitude of the radial component normalized by the compression speed.
    The white form is the foam, and the black rectangle is the impactor plate.
    (a) Slow compression, $\mathrm{v_c}=1$~mm/s and
    (b) high speed compression, $\mathrm{v_c}=100$~mm/s. 
    }
    \label{fgr:fig4}
\end{figure*}

We present exemplary fluid PIV and foam tracking results in Fig. \ref{fgr:fig4} for an STF-foam sample at low and high compression rates.
In the low-speed case shown in Fig. \ref{fgr:fig4}a, the foam collapses locally and buckles inward as the STF flows outward radially.
We observe this behavior in low-speed compressions of both foams filled with oil and with STF.
In contrast, the foam bulges out significantly when compressed quickly.
For $\mathrm{v_c}=100$~mm/s, this radial expansion begins at very low strain and continues throughout the compression. 
Once the impactor stops moving, the foam relaxes back to its original shape.
This general behavior is observed in both the STF-foam and the oil-foam composites.
In Fig.~\ref{fgr:fig4}, the magnitudes of the local flow field vectors are normalized by the compression speed $\mathrm{v_c}$. 
Comparison of panels A and B shows that, for a given strain $\epsilon$, the normalized high- and low-speed flow fields appear similar.
The area of strongest flow is concentrated near pronounced curvature in the foam boundary, due to either inward buckling or  radial expansion as a consequence of fluid drag on the foam network.

\begin{figure*}[!t]
    \centering
    \includegraphics[width=\linewidth]{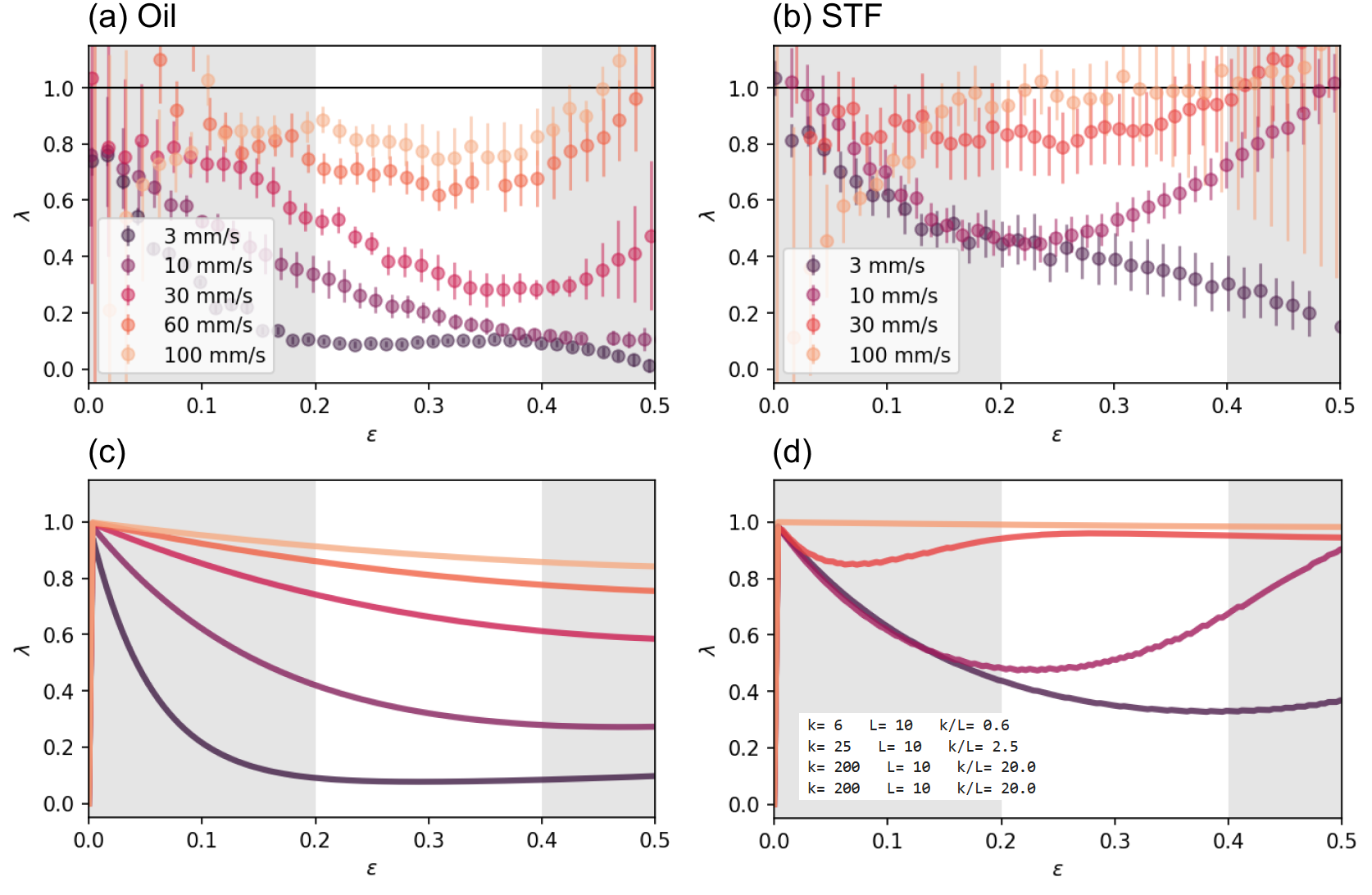}
    \caption{
    Ratio of foam and fluid radial speeds, $\lambda=\mathrm{v_{r,foam}}/\mathrm{v_{r,fluid}}$, as a function of strain $\epsilon$ for the small pore foam submerged in Newtonian (left) and STF (right) fluids.
    Experimental measurements are shown in the first row (see $\S~3.2$), and model curves (see $\S~3.3$) in the second row.
    (a)~Experimental measurements of $\lambda(\epsilon)$ for an oil-filled foam ($\eta=10$~Pa$\cdot$s).
    (b) Experimental measurements of $\lambda(\epsilon)$ for an STF-filled foam.
    (c) $\lambda$ predictions from the model for the oil-filled foam, using $k/L=20$~Pa.
    (d) $\lambda$ predictions from the model for the STF-foam, using $k/L=0.6,2.5,20,20$~Pa for $\mathrm{v_c}=3,10,30,100$~mm/s, respectively.
    Additional parameters describing the STF-foam interaction are $w=0.5 \langle d_0 \rangle$ and $\langle d_0 \rangle = 0.19$~mm.
    In all panels, the gray shaded regions indicate parts of the compression cycle that, in the experiments, may be affected by the acceleration of the impact plate; this also is where the experimental uncertainties are largest.
    In panels A and B, the data are averages of two independent measurements and the error bars correspond to one standard deviation.
    In panels C and D, the model parameter characterizing the inertial response is taken as $L/m=100$~m/kg for all $\mathrm{v_c}$.
    }
    \label{fgr:fig5}
\end{figure*}

In order to quantify the effect of drag from the suspension on the foam struts, we calculate the ratio of radial speed of the foam boundary to radial speed of the fluid directly outside the foam, $\lambda = \mathrm{v_{r,foam}/v_{r,fluid}}$. 
To this end, the foam boundary in Fig.~\ref{fgr:fig4} is subdivided along the vertical $z$-direction into 20 equal length segments at every strain step.
We then extract $v_\mathrm{r,foam}$ and $v_\mathrm{r,fluid}$ for each segment and calculate the $\lambda$ ratio as a function of compressive strain.
Figures~\ref{fgr:fig5}a and b show $\lambda$ averaged over a central subset of ten segments, chosen to include the regions of strongest flow along the foam boundary. 
The error bars indicate one standard deviation.

In the following analysis of $\lambda(\epsilon)$, we exclude strains where the impactor was either accelerating to reach the preset speed or the foam was becoming too compressed to allow for a straightforward extraction of $\lambda$ (note the large deformations of the foam boundary at $\epsilon \approx$ 60\% in Fig.~\ref{fgr:fig4} and the large error bars in Figs.~\ref{fgr:fig5}a and b at large strains and high compression speeds, where $\lambda$ also increases because the impactor must come to a stop).
These excluded regions are shaded gray in Fig.~\ref{fgr:fig5}, leaving the strain interval $0.2 \leq \epsilon \leq 0.4$ for more detailed comparison with a simple model in the next section.
Nevertheless, even cursory inspection of the curves in Figs.~\ref{fgr:fig5}a and b indicates that in all cases, except for the slowest compression of the oil-filled foam, a steady-state of the radial foam expansion is not reached.
Under steady-state conditions, the balance of foam elasticity and viscous drag from the fluid implies $\mathrm{v_{r,foam}} \approx 0$ and we thus expect $\lambda \rightarrow 0$.
Values of $\lambda$ closer to unity at a given strain therefore signal that the fluid drag is (still) causing significant radial foam expansion.
In particular, when the filler material is highly viscous, it cannot escape the foam  without pulling the network structure along with it, such that $\lambda \rightarrow 1$.

For the slowest oil-foam compression shown in Fig. \ref{fgr:fig5}a, $\lambda_\mathrm{oil}$ decays quickly and is near zero in the region of interest $\epsilon\in ( 0.2,\,0.4)$, indicating that the foam is not expanding radially.
When $\mathrm{v_c}$ is increased, $\lambda_\mathrm{oil}$ starts at a non-zero value and decreases with increasing strain.
For the two highest speeds, 60 and 100~mm/s, $\lambda_\mathrm{oil} \approx 0.8$ and stays relatively constant as a function of strain in the region of interest.
These data are shown to demonstrate the relationship between elasticity and viscous drag in foam filled with a high viscosity Newtonian fluid, resulting in a monotonically evolving $\lambda_\text{oil}$ as a function of strain and compression rate.

Performing the same measurements on the STF-foam system reveals key differences compared to the oil-foam.
First, at the slowest compression speed shown, $\mathrm{v_c}=3$~mm/s, where we do not expect significant shear thickening, the ratio $\lambda_\text{STF}$ decreases with strain in a manner similar to the 10~mm/s Newtonian oil case. 
This is surprising at first glance, as the oil viscosity is 10 times greater than that of the unthickened STF; one might therefore have expected the STF to relax faster than the oil to $\lambda$ values around zero
(but see below for further discussion).
Second, at a compression speed of 10~mm/s, the STF-foam begins to bulge radially around $\epsilon=0.25$, leading to a pronounced increase in $\lambda_\text{STF}$.

Such a change in slope of $\lambda(\epsilon)$ is unique to the STF-foam system; we never observed this in the oil-foam case.
The smooth, continuous increase in $\lambda_\mathrm{STF}$ beyond $\epsilon=0.25$ mimics what we observe for the stress response in Figs.~\ref{fgr:fig1}b and \ref{fgr:fig2}, which we related to the broad distribution of pore sizes in the foam network:
If the foam were completely uniform in its porosity, we would expect an abrupt spike in $\lambda$ upon the onset of DST as the bulk of STF throughout the foam shear thickens at once.
The fact that this upturn in $\lambda_\mathrm{STF}$ occurs at $\mathrm{v_c}=10$~mm/s agrees well with the transition from low to high dissipated energy from Fig. \ref{fgr:fig3} for the same small-pore foam filled with STF.
The change of slope in $\lambda$ for $\mathrm{v_c}=10$~mm/s can therefore be attributed to the onset of DST inside the foam once the strained pores are small enough for the effective internal shear rate to exceed $\dot{\gamma}_\mathrm{DST}$, as opposed to the 30 and 100~mm/s cases, where DST appears to take place from the very beginning of the compression.
Indeed, at speeds $\mathrm{v_c}>10$~mm/s, $\lambda_\mathrm{STF}$ increases further and  throughout the whole strain interval of interest maintains a large value, which eventually saturates at $\lambda_\text{STF}\simeq 1$ for $v_c =100$~mm/s.

\subsection{Modeling the radial foam expansion}
In order to link the radial expansion during the compression of an STF-filled open-cell foam to the distribution in pore sizes, we model the radial expansion as simply as possible.
We approximate the foam as a uniform cylinder of elastic material that is expanded radially from within by radial viscous drag. 
This effectively reduces to a one-dimensional model of a mass $m$ connected to a spring of stiffness $k$, driven by a fluid of radial speed $\mathrm{v_{fluid}}$ and  viscosity $\eta(\epsilon)$.
In the case of the STF, $\eta(\epsilon)$ depends on the strain-dependent effective shear rate. 
We assume the viscous drag to be directly proportional to the relative speed between foam and fluid, as we expect the flow inside the foam to be laminar (for a cell size of $0.85$~mm, a minimum viscosity of $\sim 1$~Pa$\cdot$s, and a maximum speed of $150$~mm/s, the Reynolds number in this regime is less than 0.2).
The equation of motion governing the radial expansion dynamics is then given by
\begin{equation}
    m\ddot{x}-\alpha(\mathrm{v_{fluid}}-\dot{x})+kx = 0.
    \label{eqn:eqn3}
\end{equation}
\noindent Here, $x(t)=r(t)-r_\mathrm{0}$ is the radial deviation of the foam edge from its unstrained value.
We define the viscous drag coefficient $\alpha = L\eta(\epsilon)$ such that $L$ has units of length and reflects the impact of foam geometry on the drag force.
In the following, $L$ will be assumed constant for a given foam.
Solutions to eqn.~(\ref{eqn:eqn3}) then give $\dot{x} = \mathrm{v_{foam}}$, which together with $\mathrm{v_{fluid}}$, enable us to calculate $\lambda$.

As the foam is compressed by the impactor moving axially at constant speed, the speed of the outflowing incompressible fluid increases.
We model the exit flow to be that out of a cylinder of initial, unstrained height $h_\mathrm{0},$ and fixed radius $r_\mathrm{0}$.
As the plate moves downwards at speed $\mathrm{v}_c$, the exiting fluid speed then reads $\mathrm{v_{fluid}(t) = v_{c }} r_\mathrm{0}/(h_\mathrm{0} - \mathrm{v}_c t)$.
Comparison with PIV data (as in Fig.~\ref{fgr:fig4}) validates this approximation over the strain interval $\epsilon\in ( 0.2,\,0.4)$. 

We numerically solve eqn.~(\ref{eqn:eqn3}) to find the best value of $k/L$ to mimic the experimental data presented in Fig. \ref{fgr:fig5}a and b in the region $\epsilon\in ( 0.2,\,0.4)$.
Inertial effects represented by the first term only affect the curvature of $\lambda$ for extremely small strain values and are negligible in the strain region of interest.
For a given fluid speed and prescribed viscosity, the solution of eqn.~(\ref{eqn:eqn3}) is thus parameterized by the ratio $k/\alpha = k/L\eta$ which compares the relative strengths of foam stiffness and hydrodynamic drag, and expresses a characteristic relaxation rate of the composite material.
In eqn.~(\ref{eqn:eqn3}), the elastic restoring force of the spring dominates over fluid drag when $kx \gtrsim \alpha \mathrm{v_{fluid}}$, which is equivalent to $k/\alpha \gtrsim \frac{\mathrm{v}_c}{h_0} \frac{r_{0}}{x}$ when rewritten in terms of $\mathrm{v_c},\,r_{0}$ and $h_\mathrm{0}$. 
When this condition is met at some $x$, further foam expansion is hindered and $\lambda \simeq 0$.
Conversely, as long as $kx \ll \alpha v_\mathrm{fluid}$, the viscous drag force dominates over foam elasticity and causes the foam to keep expanding radially such that $\lambda$ maintains a non-zero value.

In the simplest, Newtonian case, i.e. the foam filled with silicone oil shown in Fig.~\ref{fgr:fig5}c, the viscosity is $\eta = \eta_0 = 10$~Pa$\cdot$s.
In Fig.~\ref{fgr:fig5}c, a fixed ratio $k/\alpha = 2\,\mathrm{s}^{-1}$ captures the behavior of the $\lambda$ curves from experiments for all speeds shown in Fig.~\ref{fgr:fig5}a.
For $k/\alpha=2\,\mathrm{s}^{-1}$, the model predicts that the foam expansion behavior transitions from elasticity-dominated to drag-dominated at a foam expansion $x/r_{0}=0.1$ for compression speeds $\mathrm{v_c} \gtrsim 5$~mm/s.
This transition is indeed seen in the experimental data in Fig.~\ref{fgr:fig5}a, where $\lambda \simeq 0$ for $\mathrm{v_c} \lesssim 5$~mm/s, while
$\lambda$ is non-zero for for larger compression speeds and approaches unity for $\mathrm{v_c} \gg 5$~mm/s.

In the STF case, we take the viscosity to be:
\begin{equation}
\eta =
    \begin{cases} 
          \eta_0 \mathrm{~for~} \dot{\gamma}_\mathrm{eff} < \dot{\gamma}_\mathrm{DST},\\
          100\eta_0 \mathrm{~for~} \dot{\gamma}_\mathrm{eff} \geq \dot{\gamma}_\mathrm{DST},\\
   \end{cases} 
   \label{eqn:eqn4}
\end{equation}
\noindent with $\eta_0=1$~Pa$\cdot$s and $\dot{\gamma}_\mathrm{DST} = 30$ s$^{-1}$ according to the rheology shown in Fig. \ref{fgr:fig1}a.
Note that this form of $\eta$ assumes the existence of a secondary Newtonian plateau (not shown in Fig.~\ref{fgr:fig1}a).
The effective local shear rate $\dot{\gamma}_\mathrm{eff}$ is calculated  by considering the pore size distribution of the foam network. 
At each step of the integration of eqn.~(\ref{eqn:eqn3}), we  take $\dot{\gamma}_\mathrm{eff} = \mathrm{v_c}/[\left(1-\epsilon\right)d_0]$ to account for these effects using a pore size $d_0$ picked from the distribution of initially unstrained pores, which is scaled by the instantaneous strain $\epsilon$.
We then solve eqn.~(\ref{eqn:eqn3}) throughout the range of pore sizes, and take the global response to be the Gaussian-weighted sum of these "local" solutions.
We approximate the unstrained pore size distribution by a Gaussian of width $w=\langle d_{0}\rangle/2$ centered around an average pore size $\langle d_0 \rangle$, such that such that $d\in(\langle d_0\rangle -w,\langle d_0 \rangle+w)$.
As we did for the collapse of the $E_\mathrm{diss}$ data in Fig.~\ref{fgr:fig3}b, we again use $\langle d_0\rangle = \frac{1}{4}\langle d_\mathrm{cell}\rangle$, i.e., $\langle d_0 \rangle \approx 0.2$~${\mathrm{\mu}}$m for our small pore foam. 

In comparing the oil-filled and STF-filled foam data, it is important to consider how each liquid interacts with the PU foam material.
In particular, plasticization of the PU by the PEG leads to chemically induced swelling and softening of the foam material\cite{Renard_Wang_Yang_Xiong_Shi_Dang_2017}.
As a consequence, we find that the spring constants in the linear compression/extension regime for foams soaked in PEG are $\sim$10 times softer than when soaked in silicone oil (see SI).
In addition, the ester-based PU comprising our foams is hydrophilic.
Therefore, we can expect greater slip between the foam struts and oil than for the case of PEG-based STF \cite{Li_Liu_Liu_Gao_Cheng_Zhang_2023,Sochi_2011}.
In our model, plasticization decreases the spring constant $k$ while the reduced slip increases the drag coefficient $L$.
Taken together, this decreases the ratio $k/L$ more than tenfold compared to the oil-filled case.
Despite the fact that the STF viscosity $\eta_0$ is 10 times smaller at the slowest compression speeds, the even larger decrease in $k/L$ therefore indicates that the relaxation of the STF-filled foam, given by the rate $k/\alpha$, should be slower than for oil.
This slower relaxation is borne out by the data for 3~mm/s in Fig.~\ref{fgr:fig5}b, and captured by the model when we set $k/\alpha$ = 0.6~s$^{-1}$.

As the compression speed $\mathrm{v_c}$ is increased, DST is triggered according to eqn.~(\ref{eqn:eqn1}) in the region of the pore size distribution where sufficiently small pores lead to $\dot{\gamma}_\mathrm{eff} \gtrsim \dot{\gamma}_\mathrm{DST}$.  
For $\mathrm{v_c}=$10, 30, and 100~mm/s, reasonable fits to the data are obtained only if we increase $k/L$.
In particular, the model requires $k/L=2.5$~Pa to closely reproduce the slope change in $\lambda$ around $\epsilon = 0.25$ for $\mathrm{v_c}=$10~mm/s.
A value of 2.5~Pa is roughly a factor 4 larger than what we used to fit the behavior at lower speeds, before the occurrence of DST, where $k/L = 0.6$~Pa.
We interpret this increase in $k/L$ as a stiffening of the elastic response of the foam material, and thus as an increase in effective spring constant $k$, induced as shear-thickened fluid surrounds the plasticized foam struts. 
The behavior of $\lambda(\epsilon)$ for $\mathrm{v_c} = $ 30 and 100~mm/s can be reproduced if $k/L$ is enlarged further to 20~Pa.
This is in line with the idea that with increasing $\mathrm{v_c}$, a larger range of pore sizes has become involved in triggering DST, which then also leads to an enhanced stiffening of the foam network.

These results show how tracking the radial dynamics during compression, when analyzed through the lens of the model given by eqn.~(\ref{eqn:eqn3}), can provide information about the fluid-foam interaction.
Changes in this interaction are reflected with remarkable sensitivity through changes in the ratio $k/L$ required for matching the experimental data, despite the model’s simplicity.
In particular, while a single value for $k/L$ can capture the behavior of $\lambda(\epsilon)$ for the oil-filled foam irrespective of $\mathrm{v_c}$, this is not possible for the STF-filled foam.
Therefore, the speed-dependent increase of $k/L$ from 0.6~Pa to 2.5~Pa can be taken as a proxy for how the emergence of strong shear thickening inside the foam network feeds back onto that same network by strengthening its mechanical response. 
The other two parameters relevant for modeling the STF-filled foam, the width $w$ of the pore size distribution and its mean $\langle d_0 \rangle$, can be obtained from the measured distribution of cell sizes (Fig.~\ref{fgr:fig1}f), scaled by the factor $\beta = 0.25$. 
Reasonable adjustments to $w$ do not change the model output significantly, whereas it is sensitive to the value of $\langle d_0 \rangle$.
To illustrate the sensitivity of the model to parameter variation, we compare model output for a range values of $k/L$, $w$, and $\langle d_0 \rangle$ in the SI.
Finally, comparison of the top and bottom rows in Fig.~\ref{fgr:fig5} shows that the model captures much of the behavior of $\lambda(\epsilon)$ also during the early and late stages of the compression, ie., in the areas shaded gray.
However, especially for the faster impact speeds and $\epsilon > 0.4$ the experimental uncertainties become too large for quantitative analysis.   

\section{Conclusions}

We investigated the compressive response of an elastic foam network filled with a discontinuously shear-thickening fluid experimentally through stress measurements and PIV analysis to gain insights into the interaction between foam elasticity and non-Newtonian viscous drag.
Under compression, the foam network submerged in the STF amplifies shear fields in localized regions, leading to a gradual yet pronounced global thickening at compression rates significantly lower than those expected out of the bulk rheology of the fluid.
For three foams of differing pore sizes, the data shows that the rise in the total energy dissipated is set by the compression speed scaled by the average pore size of the foam, $\dot{\gamma}_\mathrm{eff}=\mathrm{v_c}/\langle d_\mathrm{0}\rangle$.
This choice of $\langle d_0 \rangle$ as the characteristic length scale in eqn.~(\ref{eqn:eqn1}) links the DST rheology to the foam geometry.

We also demonstrate that the STF greatly influences the deformation of the foam, just as the presence of the foam affects the behavior of the interstitial STF.
We show that the STF can exert drag forces on the solid network through which it travels by comparing the relative speeds of the expanding foam boundary and the local fluid through the speed ratio $\lambda$.
This quantity directly reflects the force balance between viscous drag and solid elasticity, and clearly distinguishes between unthickened (low speed, low viscosity) and thickened (high speed, high viscosity) states.
For compression rates that correspond to local shear rates approaching $\dot{\gamma}_\mathrm{DST}$, the behavior of $\lambda$ is more complex, reflecting the gradual spread of shear thickened regions throughout the material.
A simple 1D model of a spring driven by viscous drag successfully reproduces the behavior of $\lambda$ for Newtonian and strongly non-Newtonian rheologies using eqn.~(\ref{eqn:eqn1}) to estimate local internal shear rates.
By integrating a Gaussian distribution of pore sizes (and therefore of shear rates) into the model, the model is able to capture even the non-monotonic behavior of $\lambda$ at the onset of shear thickening.
The agreement between model and experiment $\lambda$ allows us to connect a small number of model parameters to physical foam features (the center and width of the pore size distribution) which are essential to explaining the smooth stress increase with strain, despite the abrupt thickening of the neat STF.
Thus, knowing only the foam elasticity, STF rheology, and the pore size distribution of the foam, we can model the balance between viscous drag and foam elasticity in STF-filled foams as a function of strain and compression rate.
This could aid in the design of advanced impact mitigation technologies with customizable mechanical properties by tailoring the fluid rheology together with the foam network architecture.
compression speeds by a common scaling with effective internal shear rate.

\section*{Safety and hazards}
Dry fumed silica is easily aerosolized and extremely hazardous to the respiratory system, so it must be handled inside of a fume hood. See OSHA guidelines for detailed regulations.

\section*{Conflicts of interest}
There are no conflicts to declare.

\section*{Acknowledgements}
The authors thank Lucy Nathwani and Philip Andrango for performing preliminary measurements.
S.L. is grateful to Tanvi Gandhi for her mentorship, and to Malcolm Slutzky, Ted Esposito, Qinghao Mao, and Andrés Cook for helpful discussions.
Support is acknowledged from the University of Chicago Materials Research Science and Engineering Center (MRSEC), which is funded by the National Science Foundation under award number DMR-2011854. 
Additional support was provided by the Army Research Laboratory under Cooperative Agreement Number W911NF-20-2-0044. S.L. acknowledges support from a MRSEC Graduate Student Fellowship and A.P. from a MRSEC Kadanoff-Rice Postdoctoral Fellowship.
S.A. acknowledges support from the French government program “Investissements d’Avenir” (LABEX INTERACTIFS, reference ANR-11-LABX-0017-01, and EUR INTREE, reference ANR-18-EURE-0010).

%--/Paper--
%%%REFERENCES%%%
\bibliography{references}
% \input{output.bbl}

% \begin{thebibliography}{36}
% \providecommand{\natexlab}[1]{#1}
% \providecommand{\url}[1]{\texttt{#1}}
% \expandafter\ifx\csname urlstyle\endcsname\relax
%   \providecommand{\doi}[1]{doi: #1}\else

\end{document}

\title{Supplementary Information to: \\
Shear thickening inside elastic open-cell foams under dynamic compression
}
\affiliation{James Franck Institute, University of Chicago, Chicago, Illinois 60637, USA}
\affiliation{Department of Physics, The University of Chicago, Chicago, Illinois 60637, USA}
\affiliation{Pritzker School of Molecular Engineering, University of Chicago, Chicago, Illinois 60637, USA}

\author{Samantha M. Livermore}
\affiliation{James Franck Institute, University of Chicago, Chicago, Illinois 60637, USA}
\affiliation{Department of Physics, The University of Chicago, Chicago, Illinois 60637, USA}

\author{Alice Pelosse}
\affiliation{James Franck Institute, University of Chicago, Chicago, Illinois 60637, USA}
\affiliation{Department of Physics, The University of Chicago, Chicago, Illinois 60637, USA}

\author{Michael van der Naald}
\affiliation{James Franck Institute, University of Chicago, Chicago, Illinois 60637, USA}
\affiliation{Department of Physics, The University of Chicago, Chicago, Illinois 60637, USA}

\author{Hojin Kim}
\affiliation{James Franck Institute, University of Chicago, Chicago, Illinois 60637, USA}
\affiliation{Pritzker School of Molecular Engineering, University of Chicago, Chicago, Illinois 60637, USA}

\author{Severine Atis}
\affiliation{James Franck Institute, University of Chicago, Chicago, Illinois 60637, USA}
\affiliation{Department of Physics, The University of Chicago, Chicago, Illinois 60637, USA}
\affiliation{Institut Prime, CNRS-Université de Poitiers-ISAE ENSMA, Poitiers, France}

\author{Heinrich M. Jaeger}
\affiliation{James Franck Institute, University of Chicago, Chicago, Illinois 60637, USA}
\affiliation{Department of Physics, The University of Chicago, Chicago, Illinois 60637, USA}
\date{\today}

\maketitle

\beginsupplement
\section{Stress-strain responses as a function of strain-rate}
In the main article text, we present a subset of all compression rates for the three different STF-foams. Here, we present the full data for all speeds.

\begin{figure}[h]
   \centering
   \includegraphics[width=5in]{all stress strain 16aug 1.png}
   \caption{Compressive stress-strain response for the small, medium, and large pore foams for all compression rates tested.
   Each curve corresponds to at least two individual compressions, and the colored areas correspond to the ranges of responses at each impact speed.}
   \label{fig:sm-dry}
\end{figure}

% \begin{figure}[h]
%     \centering
%     \includegraphics[width=2in]{supp figs/foam diagram.png}
%     \caption{Oil vvfrac, raw}
%     \label{fig:stf-raw}
% \end{figure}

\newpage
\section{Spring constant of foam networks saturated in different fluids}
To assess the effect of submerging polyurethane foams in fluids of different chemical properties (i.e. polyethylene glycol versus silicone oil), we perform compressive and tensile tests on wetted foams. 
We find that the effective spring constant $k$ of the foam network depends both on strain rate and also on the choice of submerging fluid. 
The effective spring constant of an oil-wetted foam was very similar to that of a completely dry foam, whereas the PEG-wetted foams were consistently softer than dry foam.

\begin{figure}[!h]
    \centering
    \includegraphics[width=5in]{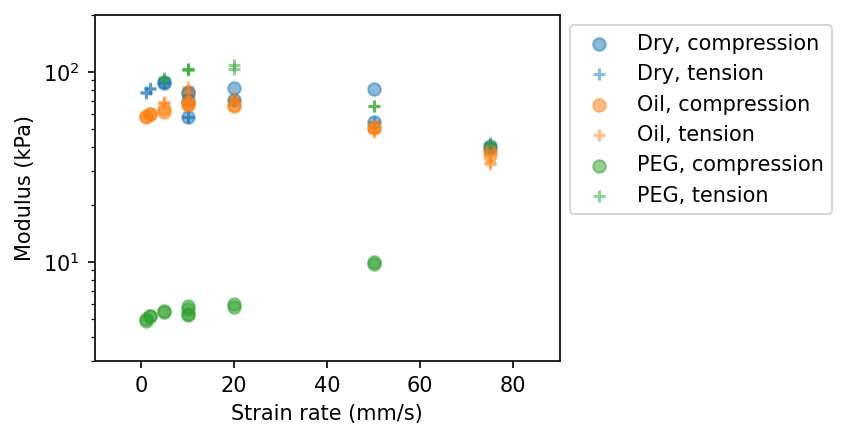}
    \caption{Measurements of the slope of the initial linear elastic regime of the compressive stress-strain response of wetted foams.
    These foams were each submerged in their respective fluids for $\sim 1$ month before being removed and wrung out for testing.
    Note how the compressive modulus of the oil-soaked foam is an order of magnitude greater on average than the PEG-soaked foam.
    Also note how different the compressive vs tensile moduli differ for the PEG-soaked foam.}
    \label{fig:springconst}
\end{figure}

\newpage
\section{Model sensitivity to input parameters}

\begin{figure}[!h]
    \centering
    \includegraphics[width=6.5in]{supp info s3.png}
    \caption{Model outputs for the small-pore foam filled with STF.
    (a) Model outputs when we vary the distribution width $w$, fix $\langle d_0 \rangle = 0.19$~mm, and $k/L=0.6,2.5,20,20$~Pa for $\mathrm{v_c}=3,10,30,100$~mm/s, respectively.
    (b) Model outputs when we vary $k/L$ and fix $\langle d_0 \rangle = 0.19$~mm and $w=0.5 \langle d_0 \rangle$.
    (c) Model outputs when we vary $\langle d_0 \rangle$, fix $w=0.5 \langle d_0 \rangle$, and take $k/L=0.6,2.5,20,20$~Pa for $\mathrm{v_c}=3,10,30,100$~mm/s, respectively.}
    \label{fig:s3}
\end{figure}

% \subsection{PIV}
% \begin{figure}[h]
%     \centering
%     \includegraphics[width=\linewidth]{supp figs/OIL vvfrac raw data 21mar2024.png}
%     \caption{Oil vvfrac, raw}
%     \label{fig:stf-raw}
% \end{figure}

% \begin{figure}[h]
%     \centering
%     \includegraphics[width=4in]{supp figs/OIL vvfrac plot 21mar2024 ALL SPEEDS.png}
%     \caption{Silicone oil vvfrac, smoothed}
%     \label{fig:si-filt}
% \end{figure}

% \begin{figure}[h]
%     \centering
%     \includegraphics[width=\linewidth]{supp figs/STF vvfrac raw data 21mar2024.png}
%     \caption{STF vvfrac, raw}
%     \label{fig:stf-raw}
% \end{figure}

% \begin{figure}[h]
%     \centering
%     \includegraphics[width=4in]{supp figs/STF vvfrac plot 21mar2024 ALL SPEEDS.png}
%     \caption{STF vvfrac, smoothed}
%     \label{fig:stf-filt}
% \end{figure}

% \begin{figure}[h]
%     \centering
%     \includegraphics[width=3in]{draft figures/si oil vs STF supp plot 6dec2023.png}
%     \caption{Caption}
%     \label{fig:si-oil}
% \end{figure}

%\clearpage
% \bibliography{dst} 
% \bibliographystyle{apsrev4-1}